\documentclass[fleqn,12pt]{wlscirep}

\newcommand{\overbar}[1]{\mkern 2.mu\overline{\mkern-2.mu#1\mkern-2.mu}\mkern 2.mu}
\newcommand*\mean[1]{\overbar{#1}}
\newcommand{\wcm}{\,\mathrm{W/cm}^2}
\title{Capturing Structural Dynamics in Crystalline Silicon Using Chirped Electrons from a Laser Wakefield Accelerator}
\author[1]{Z.-H He}
\author[2]{B. Beaurepaire}
\author[1]{J.~A.~Nees}
\author[2]{G. Gall\'e}
\author[3]{S.~A. Scott}
\author[3]{J.~R. S\'anchez P\'erez}
\author[3]{M.~G. Lagally}
\author[1]{K.~Krushelnick}
\author[1,4]{A.~G.~R.~Thomas}
\author[2]{J. Faure}

\affil[1]{Center for Ultrafast Optical Science, University of Michigan, Ann Arbor, MI 48109-2099 USA}
\affil[2]{LOA, ENSTA ParisTech, CNRS, \'Ecole polytechnique, Universit\'e Paris-Saclay, Palaiseau, France}
\affil[3]{University of Wisconsin-Madison, Madison, Wisconsin 53706, USA}
\affil[4]{Department of Physics, Lancaster University, Bailrigg, Lancashire, LA1 4YB, UK}
\affil[*]{Correspondence to Z.-H He (zhhe@umich.edu) or J. Faure (jerome.faure@ensta-paristech.fr)}

\begin{abstract}
Recent progress in laser wakefield acceleration has led to the emergence of a new generation of electron and X-ray sources that may have enormous benefits for ultrafast science. These novel sources promise to become indispensable tools for the investigation of  structural dynamics on the femtosecond time scale, with spatial resolution on the atomic scale.  Here, we demonstrate the use of laser-wakefield-accelerated electron bunches for time-resolved electron diffraction measurements of the structural dynamics of single-crystal silicon nano-membranes pumped by an ultrafast laser pulse.  In our proof-of-concept study, we resolve the silicon lattice dynamics on a picosecond time scale by deflecting the momentum-time correlated electrons in the diffraction peaks with a static magnetic field to obtain the time-dependent diffraction efficiency. Further improvements may lead to femtosecond temporal resolution, with negligible pump-probe jitter being possible with future laser-wakefield-accelerator  ultrafast-electron-diffraction schemes. 
\end{abstract}

\begin{document}
\flushbottom
\maketitle
\thispagestyle{empty}
After decades of development, electron diffraction and microscopy have become  robust and powerful techniques for determining material structure at equilibrium with atomic-scale resolution. Following the availability of femtosecond lasers in the 1980s, a number of ultrafast time-resolved techniques have been developed to study the dynamics of atomic, molecular, and condensed systems on femto- to picosecond time scales. Combining ultrafast temporal resolution with ultrahigh spatial resolution, short pulses of X-rays or electrons have the unique capability for direct structural visualisation at the atomic-level. Ultrafast X-ray pulses with durations on the order of 100 fs can currently be produced at large-scale accelerator facilities such as synchrotrons using slicing schemes~\cite{schoenlein_SCI_2000,Khan_PRL_2006} or through the use of free-electron lasers~\cite{XFEL_NatPhon}. On the other hand, producing electron pulses can be achieved using a relatively compact setup at university-scale laboratories. Compared to X-ray photons, electrons have a much higher elastic scattering cross-section and have a sub-atomic de Broglie wavelength, which makes them ideal probes for thin solid samples, gas phase, and nanoscale systems~\cite{Zewail_06,FED_2011}.

Conventional ultrafast electron diffraction techniques use DC photoguns to accelerate electrons to non-relativistic energies in the 30--100 keV range, where the space-charge effect (Coulomb repulsion) has been a major limitation to the ultimate temporal resolution. In the past decade, research toward increasing the temporal resolution has been very active, with the motivation of being able to observe the fastest structural changes that can occur in materials. Several methods have been developed, including ultra-compact electron gun designs~\cite{Waldecker_JAP_2015}, space-charge free bunches using single- or few-electron-pulses~\cite{Baum_ChemPhys_2013,Baum_2014,gliserin2015nc} and radio-frequency (RF) cavities for recompressing the electron bunches~\cite{Oudh_PRL_2010,Chatelain_APL_2012,Gao_OE_2012}. RF guns producing MeV electrons have also been used successfully for ultrafast electron diffraction~\cite{Hastings_APL_2006,Musumeci_APL_2010,murooka_APL_2011}, with electron bunches containing more charge thanks to the reduction of space-charge at higher energy. However, despite this progress, reaching resolution shorter than 100~fs has been difficult, either because of the jitter of the RF source~\cite{Chatelain_APL_2012,Gao_OE_2012,Musumeci_APL_2010} or because of the intrinsic limitations of the photoemission process at the photocathode~\cite{Baum_ChemPhys_2013}.

Recent advances in high-intensity laser development have opened up a new range of laser-plasma based X-ray and electron sources. X-rays produced in solid-density plasmas~\cite{Murnane_Sci_1991} have been widely used for time-resolved X-ray diffraction with sub-picosecond resolution for almost two decades~\cite{Rischel_Nat_1997}. Laser intensities exceeding 10$^{18}\wcm$ can now be routinely delivered by table-top laser systems, enabling remarkable progress in \emph{Laser Wakefield Acceleration} of electron bunches with energies ranging up to multi-GeV~\cite{Esarey_RMP,Malka_POP_2012}. In a laser wakefield accelerator, an ultra-intense laser pulse is focused into a plasma and excites a large amplitude plasma wave, the so-called wakefield, in which electrons can be trapped and accelerated to high energies. This technique provides acceleration gradients orders of magnitude higher than their conventional counterparts based on RF technologies, so that electrons reach MeV energies in tens of microns only, therefore providing very compact sources of electrons and also photons \cite{Corde_RMP}.  When the injection and acceleration processes are controlled properly, electron bunches can be generated with sub-10 fs duration~\cite{Lundh_NP_2011} that are intrinsically synchronised to the femtosecond laser system. Laser wakefield accelerators consequently also have great potential as all-optical sources for near jitter-free time-resolved measurements \cite{Schumaker_13}. In particular, ultrafast electron diffraction measurements using laser wakefield accelerated electrons could benefit from all these advantages.

In our study, we make use of the ultrashort bunch duration of laser wakefield accelerated electrons to perform time-resolved electron diffraction measurements of the structural dynamics of single-crystal silicon nano-membranes in an all-optical pump-probe arrangement. While solid density plasmas~\cite{Tokita_APL_2009,Tokita_PRL_2010} or  direct laser acceleration using radially polarised light~\cite{Marceau_PRL_2013,Marceau_JPB_2015} have  been proposed as sources for ultrafast electron diffraction, to our knowledge no  time-resolved electron diffraction experiments have previously been performed using a laser-plasma source. Our measurements were made possible by the development of a laser wakefield accelerator operating at kHz repetition rate with superior stability, delivering electrons in the 100 keV range~\cite{He_NJP_2013,Beaurepaire_PRX_2015}, with the transverse coherence required for electron diffraction~\cite{He_APL_2013} along with dramatic beam quality improvements enabled by active feedback control \cite{he2015coherent}. 

Using a laser wakefield accelerator has the  advantage of achieving a near jitter-free pump-probe synchronisation and producing electron bunches with uncorrelated initial durations at the few-femtosecond level. In our experiment, the emitted electrons have a significant energy spread ($\Delta E/E$=10--20\%), which results in significant bunch elongation during propagation. However, as the beam drifts in free space, it develops a well defined linear energy chirp in the longitudinal phase-space, i.e., a momentum-position correlation. Therefore, in this work we performed the time-resolved measurements with the elongated electron pulse, which is diffracted off the sample and subsequently streaked by a static magnetic field, thereby transforming temporal information to a spatial diagnostic~\cite{Mourou_APL_1982,Li_RSI_2010,Musumeci_JAP_2010,Eich_APL_2013}. This mode of operation allows experimentalists to capture the entire time history in a single image without the need to scan different time delays as in the traditional stroboscopic mode. Although our measurements have a diagnostic-limited picosecond resolution, it is clear that 100s or even 10s femtosecond resolution may be possible with further development.
 
\section*{Results}
\subsection*{Experimental set-up}
\begin{figure}[h!]
\centering
\includegraphics[width=\linewidth]{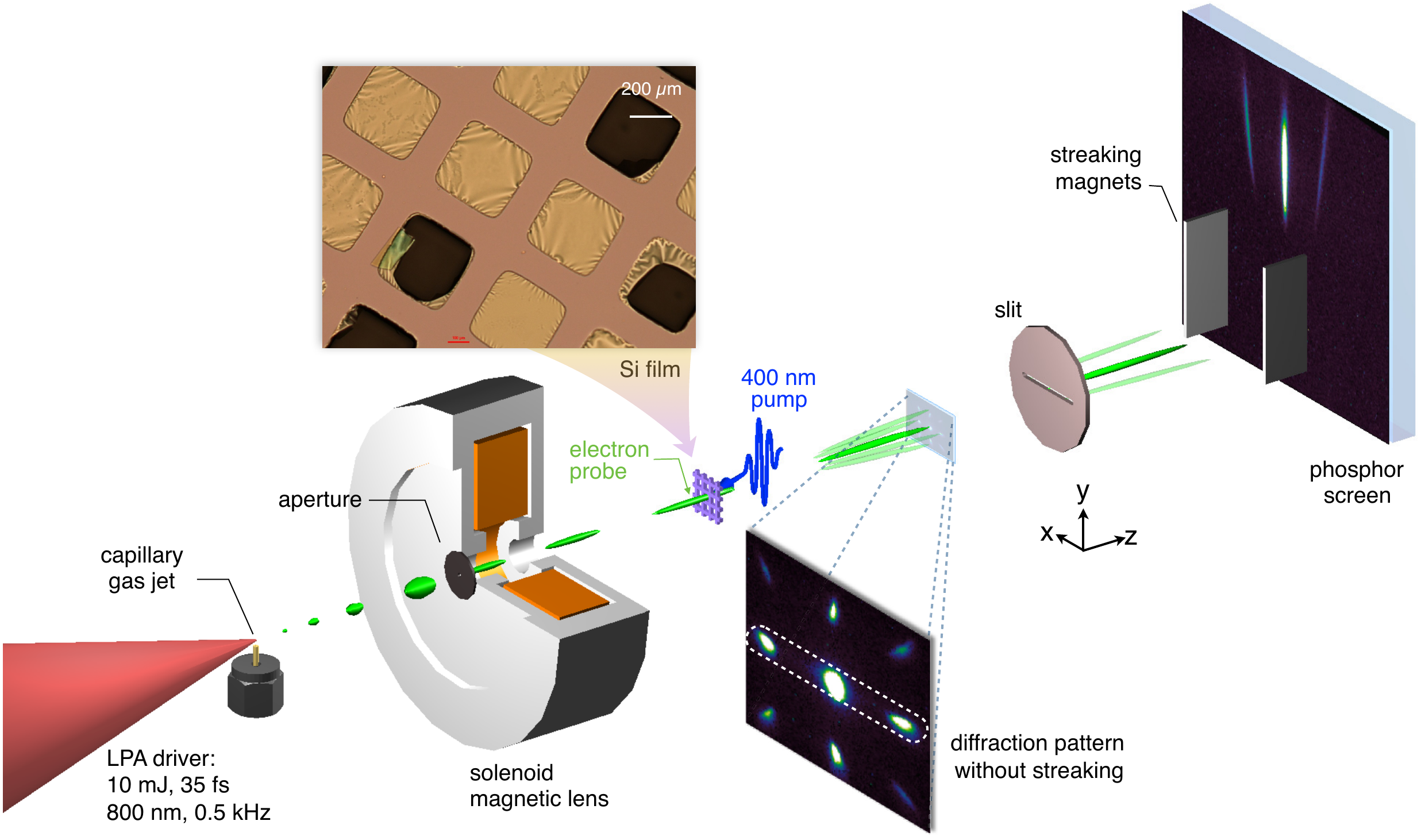}
\caption{\textbf{Experimental layout of streaked time-resolved electron diffraction} A 15-mJ Ti:Sapphire laser system is used for generating both the electron probe and the optical pump on the sample. Approximately 10~mJ of the 800~nm laser pulse energy is focussed into an argon gas jet produced by a 100~$\mu$m capillary nozzle for generating bursts of electrons. The remaining fraction of the beam is frequency doubled to 400~nm and delivered to the sample for optical excitation at an absorbed fluence of 1--2~mJ/cm$^2$. The electron beam is filtered by a 280~$\mu$m aperture before entering a solenoidal magnetic lens. A $30$~nm thick single-crystal silicon sample is placed at $d=13.5$~cm from the electron source. An optical micrograph of the array of si nano-membranes and supporting grid is displayed. The diffracted electrons enter a horizontal slit before they are spatially dispersed onto a detector screen via a pair of dipole magnets.}
\label{fig:setup}
\end{figure}

Fig.~\ref{fig:setup} shows a schematic of the experimental setup. Laser pulses with 35 fs duration at full-width-at-half-maximum (FWHM) and energies of 15~mJ were focussed to intensities of $2-3\times10^{18}\wcm$ (see Methods) into a continuously flowing gas target to generate electron bursts having energies around 100 keV at 0.5 kHz via laser wakefield acceleration\cite{He_NJP_2013}. Using a feedback loop to control the laser wavefront~\cite{he2015coherent}, the electron beam was optimised in terms of divergence and beam charge such that the total charge reached 50~fC (about 300~000 electrons). Only a  fraction (between 10--15 per cent) of the electrons were delivered to the sample after being emittance-filtered by a pinhole and collimated with a solenoid magnetic lens. The electron bunch measured  at 35 cm from the gas target had a spot size of 0.5~mm vertically and 0.7~mm horizontally at half-maximum. At the specimen plane ($d=13.5$~cm from the source), the transverse size of the electron beam was estimated to match the size of the Si samples, i.e. less than $350\;\mu \mathrm{m}$ (see Methods). The chirped bunch duration at the sample position can be estimated to be $\tau_{chirp}\simeq (d/2v)\Delta E/E$, giving $\tau_{chirp}\simeq 80$~ps for a 100~keV electron bunch with a  20\% energy spread. For streaking the electron bunch, a horizontal slit was used to select the zero-order and the (220) spots diffracted in the horizontal plane. Electrons in the selected peaks then entered a magnetic deflection region and were streaked vertically along the $y-$axis by the static magnetic field (see Fig.~\ref{fig:setup}).

For optical excitation, the 30~nm single-crystal silicon sample was pumped by 400~nm s-polarised pulses at an incident angle of $\theta=20^\circ$. The pump beam size was set slightly larger than the electron probed area to ensure spatial pump-probe overlap. The absorbed pump fluence in our experiment was  in the range 1--2~mJ cm$^{-2}$.

\subsection*{Temporal characterisation of the laser-plasma accelerated source}
\begin{figure}[h!]
\includegraphics[width=0.5\linewidth]{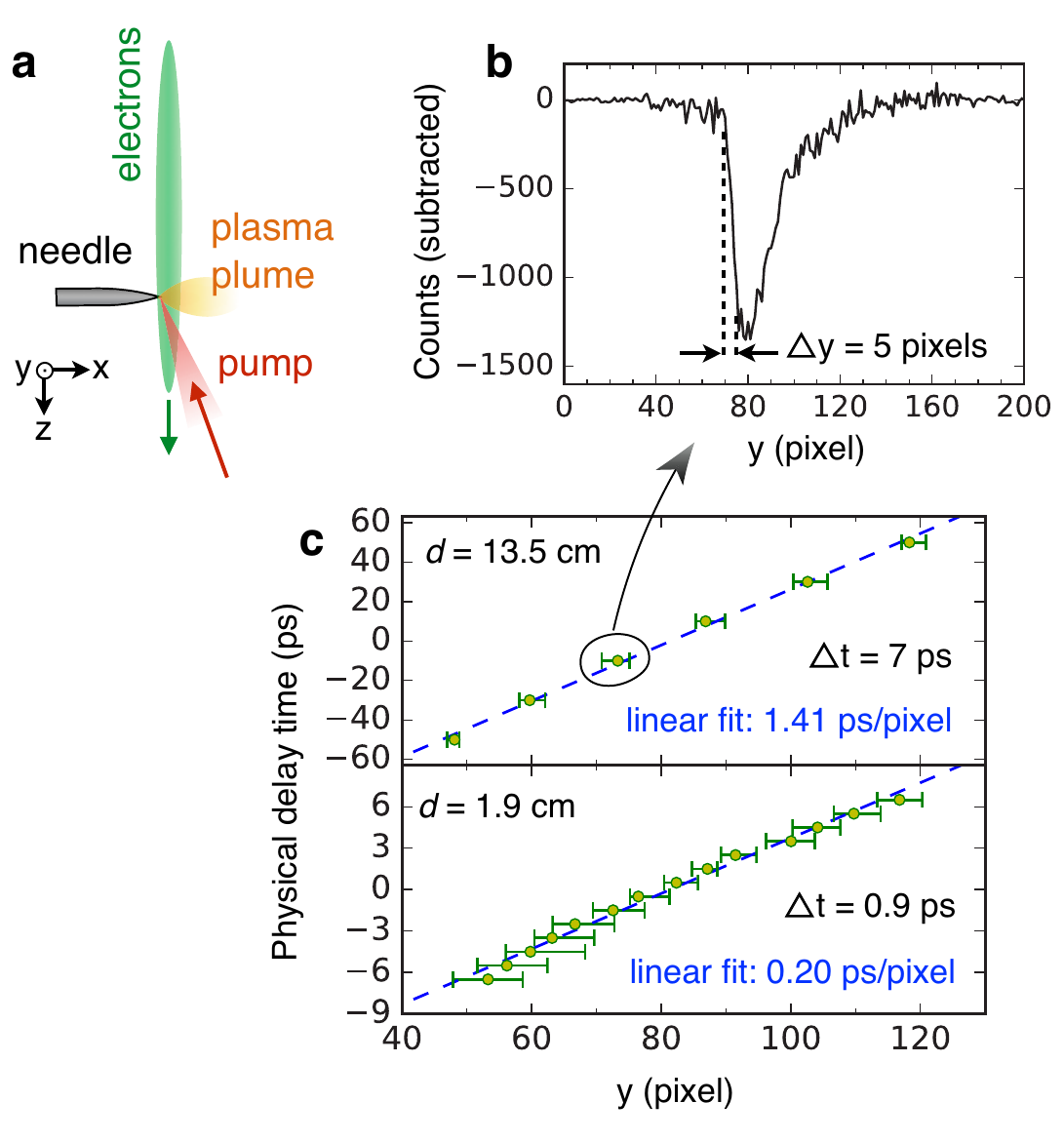}
\caption{\textbf{Delay time calibration using a tungsten needle target} (a) Schematic of experimental setup for the calibration of the pump-probe delay. (b) Intensity lineout profile $I_1-I_0$ along $y$ axis (streaking direction) showing the effect of the pump pulse on the needle. (c) Delay time plotted as a function of the position of the signal measured at location $d$~=~13.5~cm and $d$~=~1.9~cm. Horizontal error bars represent the signal transition width from 10\% to 90\% of the peak.}
\label{fig:timedelay}
\end{figure}

The time delay between the optical pump and electron probe was determined by focussing the near-infrared optical beam onto the tip of a tungsten needle target at the sample plane. The pump pulse is able to field-ionize electrons from the needle surface, thereby creating charge separation fields that cause a local deflection of the electron beam and modify the beam intensity on the detector. Therefore, the arrival of the optical pump is encoded in the streaked electron beam and can be measured by subtracting the pumped ($I_0$) and unpumped pattern ($I_1$), $I_1-I_0$, see Fig.~\ref{fig:timedelay}b. By changing the position of the delay line, one can scan the arrival of the pump pulse through the streak pattern, thereby providing a direct calibration converting the spatial position of a streak to an axis of time, see Fig.~\ref{fig:timedelay}c.

Once the delay line is set, there is no further need for scanning and Fig.~\ref{fig:timedelay}c shows that the chirped electron beam provides a temporal window of $120$~ps when the sample is placed at $d=13.5$~cm. Fig.~\ref{fig:timedelay}b also shows that the decay time of the signal is 7 ps, which sets an upper limit to the temporal resolution of this particular configuration. This is limited by the magnetic-deflection diagnostic, not the inherent slice-duration of the electron beam. The  pump-probe data in our experiments were taken at $d=13.5$~cm in order to benefit from a larger temporal window and also to keep the fluence of the main laser beam onto the Si sample at an acceptably low level.

To demonstrate that this set-up is not limited by the electron source but the magnetic-deflection diagnostic, we also placed the target closer to the source at $d=1.9$~cm. By doing so, the resolution can be significantly enhanced. The lower panel of Fig.~\ref{fig:timedelay}c, which was taken with $d=1.9$~cm, shows the temporal window is only about 15~ps because the electron bunch duration is shorter at this point but the upper limit on the resolution is now hundreds of femtoseconds. 

\begin{figure}[h!]
\centering
\includegraphics[width=0.85\linewidth]{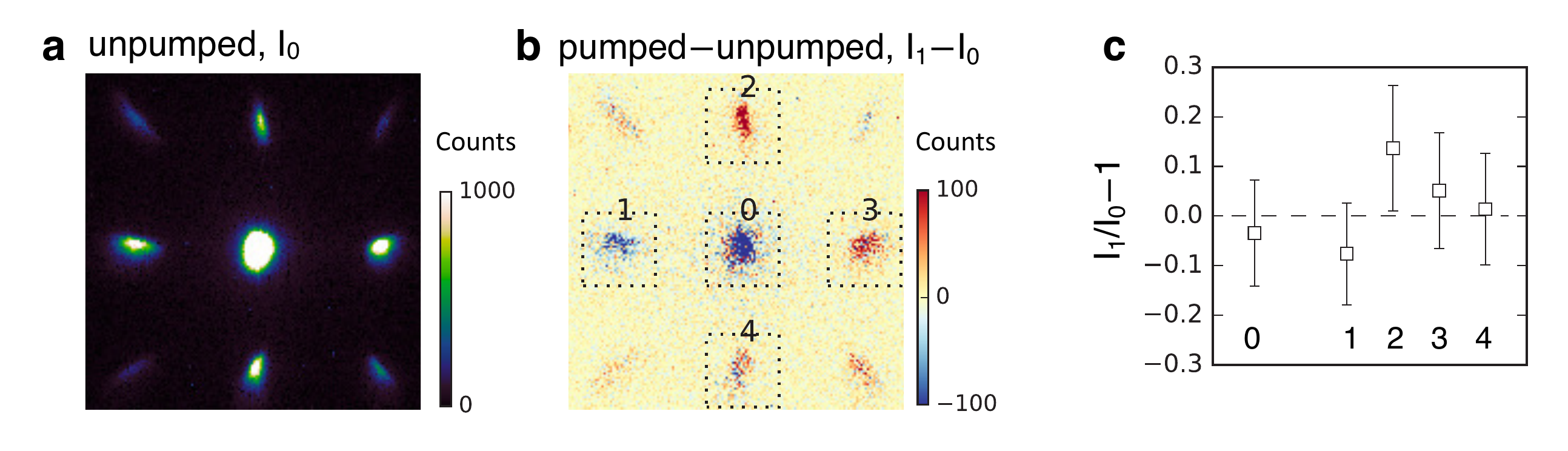}
\caption{\textbf{Diffraction pattern and pump-probe signal without streaking} (a) Diffraction pattern of single crystalline silicon, unpumped (accumulated for 200 shots). (b) Subtracted signal of the pumped and unpumped diffraction data $I_1-I_0$ averaged over 30 pump-probe cycles. The pump-probe delay is 300~ps time delay and the absorbed fluence is 1.9~mJ/cm$^2$. (c) Relative signal change for the undiffracted spot and four (220)-order Bragg spots $I_1/I_0-1$. Error bars, 95\% confidence interval.}
\label{fig:diff}
\end{figure}
\subsection*{Lattice dynamics of the silicon nano-membranes}
The quality of the single-crystal silicon membranes and their thickness of only 30~nm (see Methods) allowed us to obtain high-quality diffraction images. Figure~\ref{fig:diff}a shows a diffraction pattern of the [001] oriented Si sample. The energy spread of the electron beam leads to radial broadening of the Bragg peaks, which becomes more apparent at higher diffraction orders. Sending the pump beam causes photo-induced changes to the diffraction pattern which are easily measured by probing the sample at 300~ps after the arrival of the pump pulse. Figure~\ref{fig:diff}b was obtained by subtracting the pump diffraction pattern from the unpumped pattern. The relative changes of the central undiffracted (0-order) spot and (220) Bragg peaks are shown in Fig.~\ref{fig:diff}c. The dynamics associated with the four (220) diffraction peaks showed dissimilar trends, featuring an increase of diffraction efficiency for some (220) spots, such as spot 2 which increases by more than $10\%$ in Fig.~\ref{fig:diff}(b-c). Similar results on Si membranes were observed in Ref.~\citen{Harb_PRB_2009}. 

We may draw a few conclusions from Fig.~\ref{fig:diff}. First, because some of the peaks increase in brightness, the photo-induced response cannot be explained simply by lattice heating and the Debye-Waller effect, which always reduce the intensity of all the Bragg peaks. Second, our observation cannot be explained by a change of the structure factor, i.e., the positions of atoms in the unit cell. Indeed, it is readily seen that the structure factor for the (220) silicon peaks is maximum, i.e., each atom in the unit cell contributes to the maximum constructive interference. 

Ref.~\citen{Harb_PRB_2009} suggests that the increase of the Bragg peak intensity might originate from thermoelastic deformation of the crystal. In this scenario, electrons are first excited to high-energy states and then subsequently  relax to the bottom of the conduction band while emitting phonons and causing ultrafast heating of the Si lattice on picosecond time scales~\cite{harb06}. The key point is that the heating is non-uniform (as it follows the transverse distribution of the pump pulse), which in conjunction with the fixed boundary conditions imposed by the support grid structure, results in a thermal strain as well as surface bulging~ \cite{Harb_PRB_2009}. These effects could effectively alter the angles of the (220) planes, leading to changes in the diffracted intensity. In particular, this scenario is compatible with our experimental results as it can account for increasing Bragg peaks: a (220) peak which is initially slightly off Bragg could fully satisfy the Bragg condition after photo-excitation, explaining the increase in signal.

\begin{figure}[h!]
\centering
\includegraphics[width=0.95\linewidth]{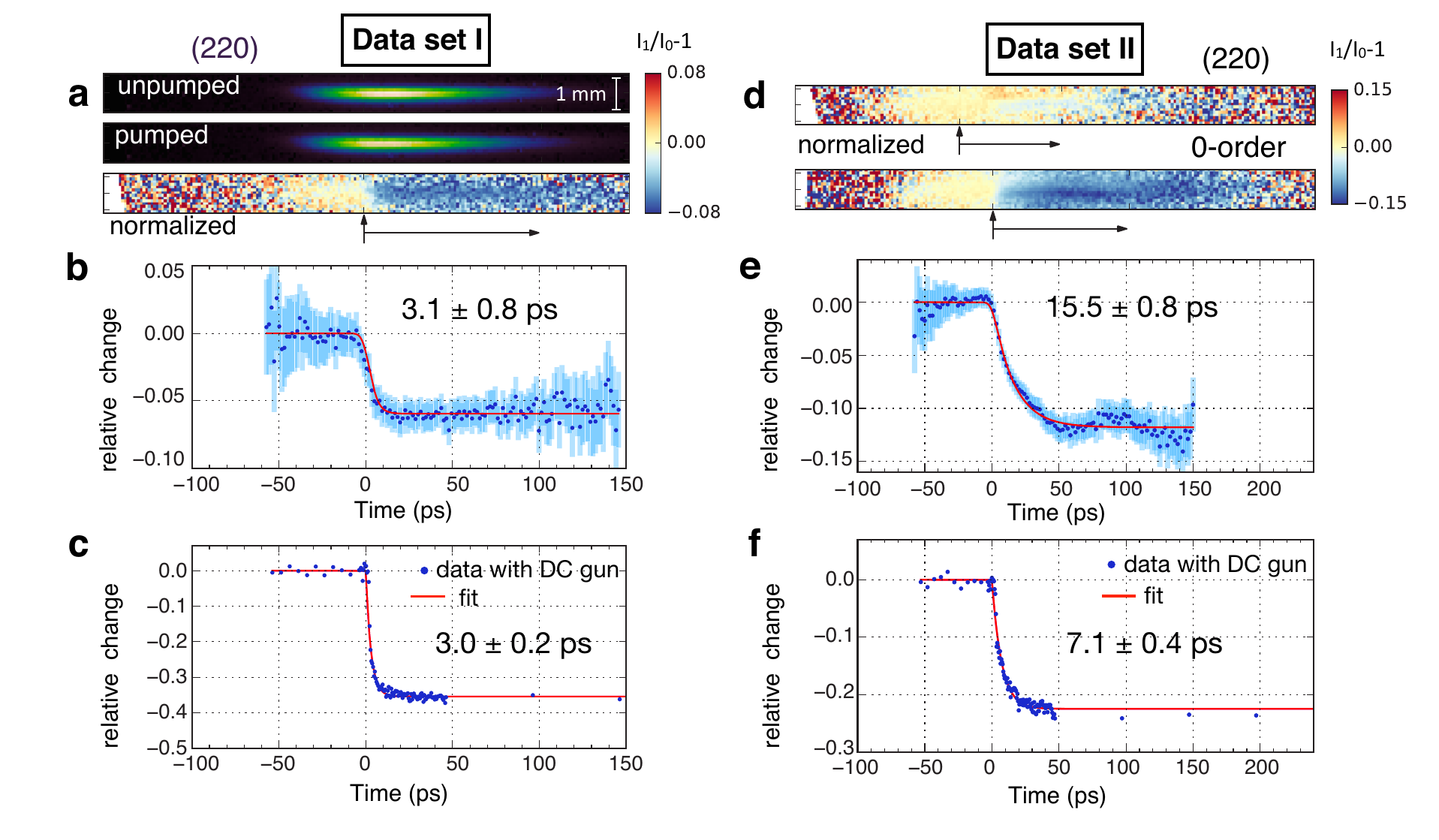}
\caption{\textbf{Time-resolved dynamics with streaked electron diffraction} Time-resolved streak signal from two different sets of data for the (220)-order and the undiffracted (0-order) spot respectively. Panel (a): raw averaged data showing the unpumped ($I_0$, upper panel) and pumped ($I_1$, middle panel) signal of the streaked (220) Bragg peak. The bottom panel represents the relative change of the diffraction peak $I_1/I_0-1$. (b) Relative change of the (220) peak obtained by transverse integration along $x$. (c) shows the (220) Bragg peak dynamics measured under similar conditions using a conventional DC electron gun. Panels (d)-(f): data from another set of measurements where the crystalline sample was tilted at a slightly different angle. (d) shows that in this case, the relative changes are larger for the 0-order spot. (e) Relative change of the 0-order. (f) Dynamics of the 0-order obtained using the conventional DC gun. Solid lines (red) are fits to the data with the decay time constants and the standard errors displayed. Vertical error bars (light blue) in (b) and (e), 95\% confidence interval.}
\label{fig:si}
\end{figure}

We now investigate the ultrafast dynamics by streaking the Bragg peaks. Results from two data sets are presented in Fig.~\ref{fig:si} for ultrafast dynamics of the undiffracted (0-order) and one of the (220)-order peaks. The pump beam profile and fluence used for both data sets were similar with an absorbed fluence of 1.3~mJ/cm$^2$. At this fluence level, the density of excited carriers is approximately $10^{21}\;\mathrm{cm}^{-3}$, i.e. 0.5\% of the valence electrons are excited to the conduction band. For data set I shown in Fig.~\ref{fig:si}(a-c), a decrease of the (220) Bragg peak intensity was observed on a few picosecond time scale, followed by a metastable state lasting more than a 100~ps. As the 0-order was saturated in this data set, we focused on the dynamics of the 0-order in data set II. In this case, no significant dynamics was observed for the selected (220) peak (Fig.~\ref{fig:si}e) while the 0-order diffraction spot shows a $\sim$12\% decrease [Fig.~\ref{fig:si}(f-g)] on a similar time scale. We attribute the different behaviour of the (220) peak in the two data sets to slightly different initial sample orientations ($<1^\circ$).

For comparison, we have also performed pump-probe experiments using a conventional DC electron gun in the stroboscopic mode, i.e., a diffraction image is collected for each pump-probe delay (see Methods). The curves for the (220) spot and 0-order dynamics are plotted in Figs.~\ref{fig:si}(c) and (f) respectively. The transients have similar features as in data sets I and II obtained with the plasma source: decay time in the ps range followed by a plateau lasting more than a 100~ps. The change in Bragg peak intensity is larger because the absorbed fluence was higher: $3.5\;\mathrm{mJ/cm}^2$. The discrepancy between the two data sets was also reproduced: data set II (\ref{fig:si}(f)) was obtained by tilting the sample by $0.8^\circ$ confirming the sensitivity of the photo-induced response to the initial orientation of the sample.

Finally, the measured data points are fitted by an exponential decay function convolved with a Gaussian instrumental response, e.g., 7~ps FWHM for the laser-plasma source (instrument limited) $3.1\pm0.8$~ps and $3.0\pm0.2$~ps for  the conventional source. This further confirms that the laser-plasma source is able to resolve the ultrafast dynamics of the Si lattice. 

All these observations can be explained as follows: first, electron-phonon coupling governs the changes of the lattice during the first few picoseconds, explaining the few picosecond drop of the (220) Bragg peak intensity. Second, electron-hole pair recombination occurs through Auger recombination. At this high density of excited carriers, $10^{21}\;\mathrm{cm}^{-3}$, screening of the electron-hole plasma mitigates the Auger recombination rate, as seen in experiments~\cite{Schank83} and theoretical work~\cite{Yoffa81}.  Therefore, it is likely that recombination occurs on time scales longer than 100~ps, explaining the longevity of the metastable excited state, which lasts more than 100~ps in our data.

\section*{Discussion}
The previous measurements confirm that laser-plasma accelerators can deliver an electron beam with sufficient quality and stability in order to probe structural dynamics at the picosecond level. We now address the question of the limits of our streaking method. 

The overall temporal resolution can be estimated as $\tau=\sqrt{\tau_{pump}^2+\tau_{uncorr}^2+\tau_{jitter}^2+\tau_{angle}^2+\tau_{streak}^2}$, where $\tau_{pump}\simeq 150$~fs (FWHM) is the pulse duration of the pump laser pulse, $\tau_{uncorr}$ is the uncorrelated electron bunch duration at the source (simulations~\cite{He_NJP_2013,Beaurepaire_PRX_2015} predict that $\tau_{uncorr}<100$~fs), $\tau_{jitter}$ is the overall timing jitter in the laser-plasma accelerator and $\tau_{angle}$ comes from the non-collinear geometry between pump and probe: $\tau_{angle}\simeq\phi \tan\theta/c\simeq 250$~fs in our case. As the electron injection and acceleration mechanisms in plasma waves rely on a deterministic evolution of plasma dynamics~\cite{he2015coherent} rather than a stochastic process, it is unlikely to induce any significant  temporal jitter at the electron emission. Most importantly, $\tau_{streak}$ is the temporal resolution depending on the specific streaking geometry. Clearly, the last term dominates our observed temporal resolution in the experiment, which is limited by the transverse electron spot size and the geometry of the streaking, subject to the space constraint of the vacuum chamber. 

If $\delta E$ is the energy resolution at $E$ for the streaking spectrometer and consider on-axis electrons, $\tau_{streak}$ is the time-of-flight difference for electrons of $E-\delta E/2$ and $E+\delta E/2$ over a distance of $d$ from their source to the sample:
\begin{equation}
\tau_{streak}=\frac{d}{v_z}\frac{\delta E}{E}\frac{m_e^2c^4}{(E+m_ec^2)(E+2m_ec^2)}
\label{eq1}
\end{equation}
where $v_z$ is the longitudinal velocity and the term $m_e^2c^4/(E+m_ec^2)(E+2m_ec^2)$ comes from the relativistic correction factor.
Substituting in the values from our silicon pump-probe experiment ($d$~=~13.5~cm, $\delta E$~=~1.6~keV, $E$~=~90~keV) resulted in $\tau_{streak}$~=~6.0~ps (evaluated near the time zero), in agreement with the measured resolution.

There is no physical limit for reducing $\tau_{streak}$ to the same level of $\tau_{uncorr}$ and $\tau_{jitter}$, which may be on the order of 10~fs. As can be seen from equation (\ref{eq1}), further improvement on the temporal resolution is straightforward by increasing the streaking strength (i.~e., decreasing $\delta E$), at the cost of either lower signal-to-noise ratio or longer acquisition time. Another way to increase the temporal resolution is to reduce $d$, the source to sample distance. As shown before with the delay calibration using the needle target, $d$~=~1.9~cm yields a $\tau_{streak}$ of 0.84~ps, under identical streaking conditions. One potential issue with decreasing $d$ comes from the fact that the co-propagating laser radiation used to drive electron acceleration might be too intense for the sample and needs to be blocked or reflected while allowing the electron beams to pass. Finally, another way to decrease $\tau_{streak}$ is to increase the electron energy $E$ while keeping the resolution of the magnetic spectrometer the same. This may become possible in the near future with recent results showing that multi-MeV electrons can be generated~\cite{Beaurepaire_NJP_2014,Goers_PRL_2015}.

Stability of the laser-plasma electron source is the main challenge in our experiments. Shot-to-shot noise (fluctuation in the unpumped data) near the central part of a streak typically has a signal-to-noise ratio about 7 dB. Sufficient statistics from large data accumulation is necessary to reveal the observed effects of a few percent changes in the streaked diffraction signal. For example, to achieve a standard error on the 1\% level, the number of pump-probe cycles required for averaging will be 400, translating into an acquisition time of $~10$ minutes. 
Note that in streaking mode, the fluctuations in electron charge and energy spectra can be averaged out without affecting the temporal resolution.
This technique may also provide useful information on irreversible phenomena such as electronically driven phase transition of silicon~\cite{Harb_PRL_2008} in a single-shot fashion provided that a sufficiently sensitive detection system is available.

Upgrading our setup with a signal intensifier such as a micro-channel plate based detector or using a higher collection efficiency device, such as a fibre-optically coupled charged-coupled device camera could potentially enable orders of magnitudes increase in the signal-to-noise ratio. These enhancements are now within present technical capabilities so that laser-plasma accelerators could soon provide a path for sub-100-fs or even sub-10-fs ultrafast electron diffraction~\cite{faure16}.

\section*{Methods}
\subsection*{Laser system}
Laser pulses generated at 500 Hz repetition-rate with a centre wavelength of 0.8~$\mu$m and a full-width-at-half-maximum pulse duration of 35~fs, produced by the Relativistic Lambda-Cubed ($\lambda^3$) laser system, were split into two optical beams by a pellicle beam splitter. 
One of them, containing the majority of the laser-pulse energy, was focussed onto argon gas flow by an $f/2$ off-axis parabola for the generation of the electron probe bunches. The second beam was frequency-doubled in a $\beta$-barium borate (BBO) crystal and focussed by a lens of focal length 10~cm for pumping the sample. A half-wave plate was installed before the BBO crystal to allow for the attenuation of the frequency-doubled light. A 2~mm BG40 bandpass filter and a low-dispersion 400~nm dielectric mirror were used to minimise the remaining energy at the fundamental wavelength. Calculating the dispersion for the pulse broadening effects from the optical elements used in the beam path yields a pulse duration of 150~fs. During the time calibration with the tungsten needle, the BBO crystal and BG40 filter were removed, and the dielectric mirror was replaced with one designed for near-infrared wavelengths to allow irradiation of the needle target at 800~nm. The optical pump beam size at the specimen plane was varied by translating the focussing lens along its optical axis. The relative temporal delay between the electron probe and optical pump was adjusted by an optical delay stage installed in the pump beam path.

\subsection*{Electron generation and diagnostics}
Electrons were generated through the intense laser pulse interacting with a continuously flowing argon gas jet. The focussed laser pulse ionises the gas and excites plasma waves, which lead to emission of electron bunches via a process known as laser wakefield acceleration (see ref.~\citen{He_NJP_2013} for the details of the acceleration mechanism). During normal operation, the vacuum chamber pressure remains below $10^{-3}$~mbar. After passing through an aperture, the electrons are nearly collimated by a solenoidal magnetic lens.
Electrons were measured at 35~cm downstream on a high resolution scintillating screen (J6677 FOS by Hamamatsu), which was imaged using a lens coupled 14-bit electron-multiplying charged-coupled device camera (EMCCD, Andor Luca-R).
The data in Fig.~\ref{fig:si}c) and f) was obtained with a conventional DC electron gun. Electrons were generated on a gold photocathode by a frequency tripled 266 nm Ti:sapphire laser, leading to a charge of approximately $2000$ electrons per bunch at 1-kHz. Electrons were then accelerated to 45~keV by a DC field and collimated with a solenoidal lens. The Si sample was placed 40 mm after the photocathode, and the electron bunch duration was estimated to be $ 300$~fs. Time-resolved data was taken in stroboscopic mode by recording a diffraction pattern for every pump-probe delay.

\subsection*{Sample fabrication}
The Si nano-membrane samples are prepared from a silicon on insulator wafer \cite{scott07,cavallo10}, which is composed of a 78 nm Si template layer, separated from a bulk Si handle wafer by a 400~nm buried oxide layer (BOX) and capped with a 150 nm thermal oxide to protect the template layer.  An array of squares ($350\;\mu\mathrm{m}\times 350\;\mu\mathrm{m}$ each) are etched through the handle wafer with deep reactive ion etching (DRIE) using C$_4$F$_8$ and SF$_6$, which etches through to the buried oxide etch stop.  The capping oxide, and the buried oxide layer in the etched regions, are then selectively removed with hydrofluoric acid. This forms an array of Si nano-membranes supported only around their perimeter by the original silicon on insulator wafer. The entire structure is then thinned to 30 nm with CF$_4$ reactive ion etching. It is clear from the optical micrograph in Fig.~\ref{fig:setup} that some rippling exists around the perimeter of the membranes, likely a consequence of strain relaxation in the membrane after the supporting BOX is removed. Nonetheless, large areas towards the centre remain relatively flat. A nano-membrane yield of approximately $50\%$ results from both sample removal from the DRIE support and subsequent oxide stripping in hydrofluoric acid.

\subsection*{Data acquisition and analysis}
Each streaked diffraction image was taken using about 300 electron pulses (0.6~s exposure time). Every pump ``on'' state exposure ($I_1$) was immediately followed by a pump ``off'' shot ($I_0$). Alternating between the signal and reference shots alleviates long-term stability issues such as heat accumulation and signal decrease of the electron charge due to drifting laser/gas profile conditions over the course of measurement. The data acquisition was automated to record about 1000 cycles, over a period of about 30 minutes. The signal ratio between the (220)-order and zero-order, i.e., the diffraction efficiency in every unpumped shot, was monitored to ensure that the induced effects in the sample were reversible within the sampling interval. The uncertainties in the relative signal $R=I_1/I_0$ were determined from the measurement using the standard error propagation  $\sigma_R=(\mean{I_1}/\mean{I_0})\sqrt{(\sigma_{I_0}/\mean{I_0})^2+(\sigma_{I_1}/\mean{I_1})^2-2\sigma_{I_0I_1}/\mean{I_0}\mean{I_1}}$, which is shown for the 95\% confidence interval in Fig.~\ref{fig:si} and used as weights for the least-squares fitting.

\section*{Acknowledgements}
This work was supported by the US Air Force Office of Scientific Research under award number FA9550-12-1-0310, the US National Science Foundation under contract 1535628 and the US Department of Energy under contract DE-NA0002372.  Fabrication and characterisation of Si nano-membrane samples (S.A.S. and M.G.L.) was supported by the U.S. Department of Energy Grant DE-FG02-03ER46028. J. F. acknowledges the support of the European Research Council, contract No. 306708, ERC Starting Grant FEMTOELEC).

\section*{Author contributions statement}
Z.H. directed the laser wakefield experiment with the participation of B.B., J.N., J.F. and analysed the results. 
G.G. and J.F. directed the electron gun experiment and analysed the results. S.A.S., J.R.S.P. and M.G.L. prepared the Si membranes.
Z.H. and J.F. wrote the manuscript. 
J.F., A.G.R.T., K.K. and J.N. provided overall supervision for the project. 
All authors discussed the results and contributed to the manuscript.
\section*{Additional information}
\textbf{Competing financial interests:} The authors declare no competing financial interests.
\end{document}